# Hole burning in polycrystalline C60: the final answer to the long pseudocoherent tails

María Belén Franzoni and Patricia R. Levstein*

*Facultad de Matemática, Astronomía y Física,*

*LANAIS de RMS, Universidad Nacional de Córdoba,*

*Ciudad Universitaria, 5000, Córdoba, Argentina*

Jésus Raya and Jérôme Hirschinger

*Laboratoire de RMN de la Matière Condensée, Institut Le Bel,*

*Université Louis Pasteur, 67000, Strasbourg, France*

## Abstract

New NMR experiments reveal the origin of the "pseudocoherence" that leads to long tails in some Carr-Purcell-Meiboom-Gill (CPMG) sequences in contrast with the decay of a standard Hahn echo. We showed in [M.B Franzoni and P.R. Levstein, Phys. Rev. B 72, 235410 (2005)] that under certain conditions these CPMG sequences become spoiled by stimulated echoes. The presence of these echoes evidences that the long tails arise on the storage of magnetization in the direction of the external magnetic field (z-polarization), involving spin-lattice relaxation times of several seconds. Hole burning experiments confirm the presence of a highly inhomogeneous line and show that the flip-flop processes are able to homogenize the line in times agreeing with those obtained through a Hahn echo sequence. As expected, the decay of the stimulated echoes is not sensitive to the quenching of the dipolar interaction through an MREV16 sequence, while the Hahn echo decay increases by a factor of 20. CPMG sequences with hard pulses and different temporal windows clearly show the relevance of the dipolar interaction in between the pulses for the generation of the long tails. In particular, the more time the flip-flop is effective (longer interpulse separation), the shorter are the characteristic times of the long tails.

## I. INTRODUCTION

Electron and nuclear spins in the solid state are good qubit candidates in order to implement gates for quantum information processing. As a consequence, many groups are nowadays working on decoherence on several kinds of spin systems. Some of the investigations point directly to the coherence time of the nuclear spins,[1–5] while others involve decoherence of electron spins because of the presence of a fluctuating nuclear spin lattice.[6,7] Observation of very long tails in measurements of spin coherence times $T_2$ using some variations of the Carr-Purcell-Meiboom-Gill ($CPMG$) sequence[8–10] in samples as $^{29}$Si,[1] $C_{60}$,[4] $Y_2O_3$[5] created great expectation. This is because the viability of quantum devices and quantum processors relies on long coherence times to perform the engineered operations.[11–13]

Several pulse sequences, besides the standard Hahn echo[14] $\left(\frac{\pi}{2}\right)_X - \tau - (\pi)_Y - \tau - echo$ have been used to measure spin-spin relaxation times $T_2$, denoted as

$$
\begin{aligned}
CP1 &: \left(\frac{\pi}{2}\right)_X - [\tau - (\pi)_X - \tau - echo]_n \, ; \\
CP2 &: \left(\frac{\pi}{2}\right)_X - [\tau - (\pi)_X - \tau - echo - \tau - (\pi)_{-X} - \tau - echo]_n \, ; \\
CPMG1 &: \left(\frac{\pi}{2}\right)_X - [\tau - (\pi)_Y - \tau - echo]_n \, ; \\
CPMG2 &: \left(\frac{\pi}{2}\right)_X - [\tau - (\pi)_Y - \tau - echo - \tau - (\pi)_{-Y} - \tau - echo]_n
\end{aligned}
\quad (1)
$$

Indeed, we emphasize that none of these measurements of $T_2$ yields the "true decoherence time" because the dipolar evolution itself is reversible. However, as it has been reported, even when the dipolar interaction is reversed the observed decay time still depends on it.[15–18] Then, we are still very interesting in the "decoherence" times given by these spin-spin relaxation measurements.

The spin-spin relaxation time, i.e. the dipolar limited coherence time, measured for $C_{60}$ using the Hahn echo sequence was $T_{2HE} = 15$ ms and similar results were obtained for $CP1$ and $CPMG2$. On the other hand, for the $CPMG1$ and $CP2$ the decay times were in the order of seconds. Under these striking results, an effort to distinguish between coherence and "*pseudocoherence*" times must be done.

In a previous work,[4] we proved that these long times are a consequence of the formation of stimulated echoes, after two $\pi$ pulses, that are not analytically expected. In the proof, two conditions were necessary for the formation of these echoes,

1. an rf field inhomogeneity or a highly inhomogeneous line able to produce different tilting angles in different sites of the sample.

2. the absence of spin diffusion (noneffective flip-flop interactions).

In this work, we show new experiments that confirm that both conditions are satisfied in $C_{60}$, verifying our previous hypothesis.

Besides a hole burning experiment, a careful study to distinguish between coherence and pseudocoherence times obtained for the $CPMG1$ and $CP2$ sequences was made, by isolating the stimulated echoes and measuring their characteristic decay times under several conditions. Also, the efficiency of the dipolar interaction in order to avoid the stimulated echo formation was studied by allowing enough time for the operation of the flip-flop dynamics, emphasizing the importance of condition 2). Our results do not agree with the interpretation of the long decay times reported recently by Li et al.[5]. They assign the long tails to the dipolar evolution that occurs during the pulses. Indeed, one could substitute condition 1) by using very long pulses. In this case, the one flip terms produced by the dipolar interaction during the pulse are similar to those produced by the off-resonance or the $H_1$ inhomogeneity. However, condition 2) is still necessary and if the dipolar interaction is important during the pulse, it will be during the inter-pulse times unless these intervals are too short compared with the duration of the pulses. This last situation is not usual in $CPMG$ measurements in the solid state.

The key to explain the origin of the long tails requires a deep understanding of the role of the dipolar interaction during and in-between pulses. The effect during the pulses was analyzed from experiments reported by Li et al.[19] which indicate that for typical solid-state NMR conditions, $\gamma H_1/2\pi>$FWHM (full width at half maximum), the tails are independent on $H_1$. For the interpulse analysis, we applied the $MREV16$ decoupling sequence to study its effects on the stimulated echoes formation.

## II. EXPERIMENTAL METHODS AND RESULTS

The experiments were performed in a Bruker Avance II spectrometer, and in a Bruker 300-MSL operating at 75, 45 MHz for $^{13}C$ and in a Bruker Avance under a 11.75 T magnet (125, 75

MHz for $^{13}$C). The probes were Bruker high power CPMAS used under static conditions. The rotor sizes were 18 mm long and 4 mm or 3.2 mm outer diameter. All the experiments were conducted on-resonance at room temperature. Typically, for almost every experiment the $\frac{\pi}{2}$ pulse was set as 5 $\mu$s.

A commercial sample of polycrystalline $C_{60}$, 99.5% purity was used without further processing. The $T_1$ observed for the sample was $T_1 = 37$ s at the 7 T magnet and $T_1 = 24$ s at the 11.75 T one.

In the following subsections, a detailed description of the experiments we performed are provided.

## A. Hole burning in $C_{60}$

The inhomogeneity of the $C_{60}$ line is well shown through a hole burning experiment[10]. The experiment consists on two $\frac{\pi}{2}$ pulses, the first a selective one of duration $t_p$. After excitation of a portion of the line with the first pulse, free evolution is allowed during a time $\tau$ until a non selective pulse is followed by acquisition (Fig.1). If the line is inhomogeneously broadened, the portion of the line excited with the first pulse will form a hole. In contrast, an homogeneously broadened line will collapse as a whole, i.e. its magnitude will diminish but no hole will be observed.

Indeed, a measure of the inhomogeneity of a line may arise from the minimum time between pulses needed to cross from the hole formation to the collapsed line situation.

We performed the experiment in polycrystalline $C_{60}$ and observed the formation of holes, manifesting that the line is inhomogeneously broadened. The width was calculated by fitting each spectrum to a Lorentzian line and the hole width as a function of the selective pulse duration was studied. As can be seen in Fig.1 the hole width is not inversely proportional to $t_p$ as ideally expected. Indeed, an exponential relationship was obtained whose characteristic time, similar to $T_{2HE}$, indicate that the dipolar interaction is starting to be operative during the selective pulse.

For two different durations of the selective pulse $t_p$, we studied the recovery of the hole by varying the evolution time between pulses, $\tau$ (see Fig.2). The quantification of the depth of the holes is fairly straightforward because the holes do not change shape as they recover.

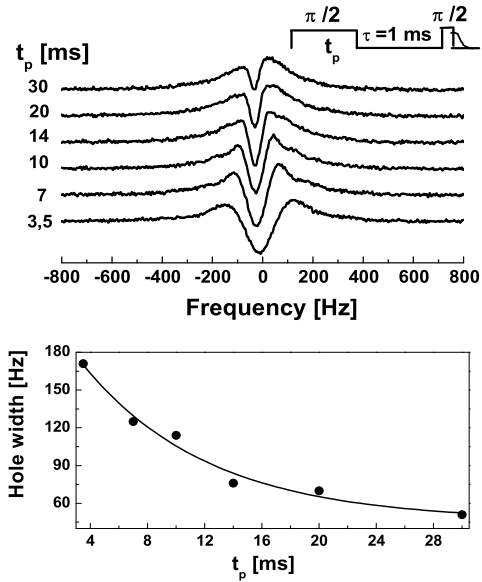

FIG. 1: Top: hole burning sequence and spectra with holes for different selective pulse durations, $t_p$. Bottom: The Lorentzian widths as function of pulse durations fit an exponential decay with characteristic time $t_c = 9ms$.

Following Kuhns and Conradi[20] to measure the hole depth, the line shape is multiplied by a rectangular function which is $+1$ in the middle of the holes (in a frequency interval as wide as the hole's full width half maximum, FWHM), $-1$ on the shoulders of the hole, and zero elsewhere. The total area of this function approaches zero; the rectangular function is slightly adjusted so that an unburned line-shape multiplied by the rectangular function has zero area. The total area of the product (experimental line shape times rectangular function) is taken as a quantitative measure of the hole depth.

As shown in Fig.2 the results obtained for $t_p = 20$ ms and $t_p = 30$ ms are almost coincident for every evolution time $\tau$. The hole area as a function of $\tau$ is well fitted to a Gaussian, yielding a characteristic decay of approximately 13.9 ms similar to the coherence time $T_{2HE}$ obtained with the Hahn echo. This is another confirmation that $T_{2HE}$ is the coherence time associated with the dipolar interaction.

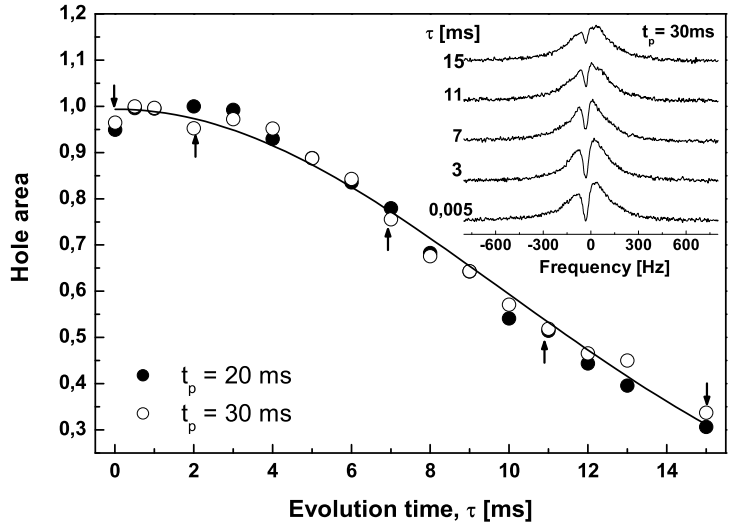

FIG. 2: Line recovery, measured through the hole area, as a function of evolution time between the pulses, for two selective pulse durations. The curve fits a Gaussian with a characteristic time of 13.9 ms. In the inset, the spectra of the intermediate times indicated by arrows in the curve, are shown.

### B. Stimulated Echo

We developed a sequence of three pulses which provides the stimulated echo decay time. As shown before,[4] in the stimulated echo ($SE$) sequence,

$$SE : \left(\frac{\pi}{2}\right)_{\varphi_1} - t_f - (\beta)_{\varphi_2} - t_v - (\beta)_{\varphi_3} - t_f - acq, \qquad (2)$$

the stimulated and the normal echoes have the same phase if the pulse phases are taken as in the $CPMG1$ ($\varphi_1 = X$; $\varphi_2 = Y$; $\varphi_3 = Y$) or as in the $CP2$ ($\varphi_1 = X$; $\varphi_2 = X$; $\varphi_3 = -X$) sequences. On the other hand, for the $CPMG2$ ($\varphi_1 = X$; $\varphi_2 = Y$; $\varphi_3 = -Y$) or the $CP1$ ($\varphi_1 = X$; $\varphi_2 = X$; $\varphi_3 = X$) sequences the echoes appear in opposite phases because the phase of the stimulated echo is inverted. Thus, by applying the $SE$ sequence (2) as in the $CPMG1$ and then as in the $CPMG2$ sequence, multiplied by 1 and $-1$ respectively (by inversion of the detection phase), we get rid of the normal echo.

By fixing $t_f = 8$ ms and varying $t_v$, we built the stimulated echo decay curve (see Fig.

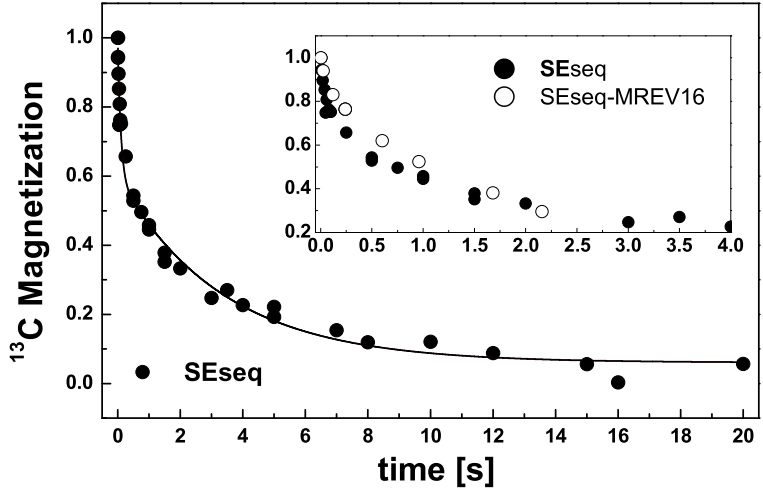

FIG. 3: Stimulated echo as a function of $t_v$, applying the three-pulse sequence with $\beta = \pi/2$ and $t_f = 8$ ms. The curve is fitted with a double exponential decay, the characteristic times obtained were $t_{s_{SE}} = 130$ ms and $t_{l_{SE}} = 4.9$ s . Inset: comparison of the stimulated echo sequence with and without $MREV16$ decoupling. No difference is observed.

3). The curve manifests two time regimes that, fitted with a double exponential decay, yield a short decay time $t_{s_{SE}} = 130$ ms and other of $t_{l_{SE}} = 4.9$ s. The long decay time is in the same order than those observed in the $CPMG1$ and $CP2$ sequences.[4]

The long time obtained for the stimulated echo ($t_{l_{SE}}$) is well understood since it brings back magnetization preserved in the z axis to the plane. As we have proved,[4] the cause of the long tails in the $CPMG1$ and the $CP2$ sequences is the constructive interference between normal and stimulated echoes. Now, we are showing that, as expected, the stimulated echoes are the ones with long decay times. We performed $SE$ sequences with $\beta = \pi$ and $\beta = \pi/2$, and both of them yielded the same decay times. In fig. 3, we show the $SE$ sequence with three $\pi/2$ pulses which gives the maximum signal to noise ratio.

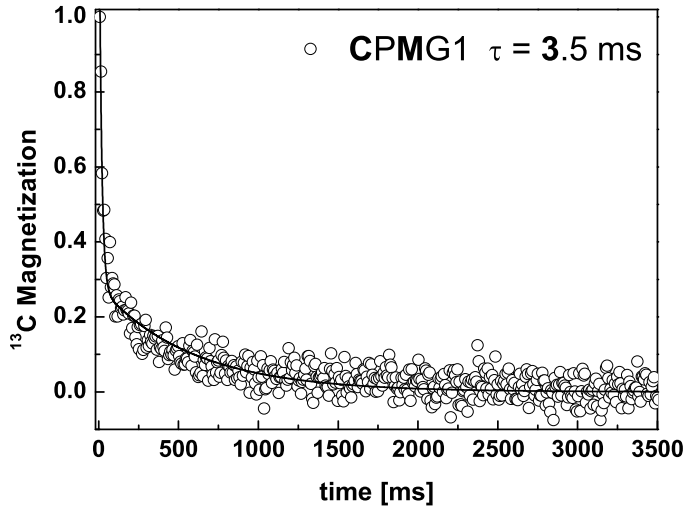

FIG. 4: $CPMG1$ experiment for $\tau = 3.5\ ms$. Double exponential decay behavior fitted with $t_s = T_{2HE}$ yields a long decay time $t_l = 570\ ms$.

### C. Long decay times vs. flip-flop dynamics.

We performed the $CPMG1$ and the $CP2$ experiments for different temporal windows, $\tau$ (in fact, $2\tau$ between the $\pi$ pulses). For each $\tau$ we observed that the magnetization shows a double exponential decay with a characteristic short time ($t_s$) and a long one ($t_l$). For all the experiments, the obtained $t_s$ was very well approximated by $T_{2HE}$, so we fixed $t_s = T_{2HE}$ and we calculated $t_l$. In Fig. 4 we show a $CPMG1$ experiment with $\tau = 3.5\ ms$ fitted with a double exponential with $t_s = 15$ ms.

After doing the same analysis for every $\tau$, we plotted $t_l$ vs $\tau$ for the $CPMG1$ and $CP2$ sequences (Fig. 5). It is noticeable that $t_l$ decays as a function of $\tau$ with a characteristic time which again resembles $T_{2HE}$!

These experiments are in very good agreement with our hypothesis and some numerical results. They show that the stimulated echoes and, consequently the long tails, are a manifestation of the absence of flip flop dynamics: a shorter $\tau$ (no flip-flop) leads to a longer decay time, $t_l$.

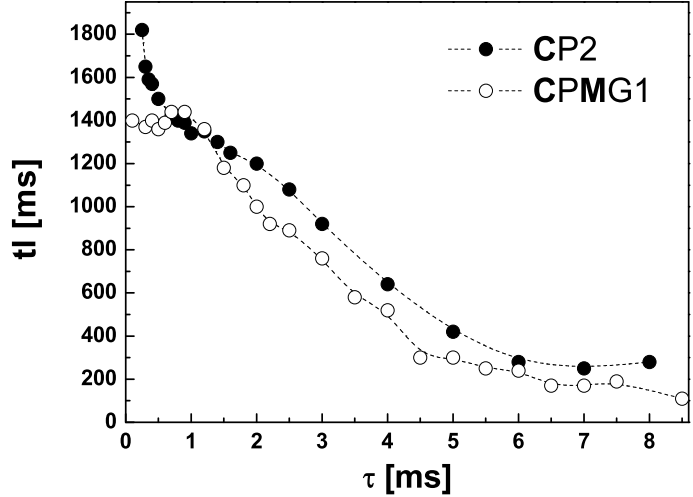

FIG. 5: The $CPMG1$ and $CP2$ experiments were repeated for many time windows $\tau$. For each train of echoes a double exponential decay was fitted, and setting one characteristic time as $T_{2HE}$, a longer one, $t_l$, was obtained. $t_l$ is plotted as a function of $\tau$.

### D. Dipolar Decoupling

The $MREV16$ sequence[21,22] consists of sixteen properly phased and separated $\pi/2$ pulses of duration $\tau_{\pi/2}$, with a minimum separation time $\tau_{MREV}$. A single $MREV16$ cycle lasts $24\tau_{MREV} + 16\tau_{\pi/2}$. As can be seen from Average Hamiltonian theory,[23] the $MREV16$ sequence averages out the zero and first order of the dipolar interaction.

We performed a $Hahn-MREV16$ echo measurement, by applying $c$ times the $MREV16$ sequence before and after the $\pi$ pulse in the Hahn echo sequence. This is, the dipolar decoupling was operative all the time after the $\pi/2$ pulse, up to the formation of the echo, where signal acquisition starts. To build the magnetization decay curve as a function of time shown in Fig.6 we used $\tau_{MREV} = 0.1$ ms and $\tau_{MREV} = 0.2$ ms and varied $c$ from 1 to 80. We obtained $T_{2HE}^{MREV} = 300$ ms, instead of $T_{2HE} = 15$ ms obtained with the standard Hahn echo.

Another revealing experiment performed to the C$_{60}$ sample was the $CPMG1-MREV16$. The $CPMG1 - MREV16$ sequence is a $CPMG1$ with a variable number ($c$) of $MREV16$ cycles applied between the $\pi$ pulses. In this case, $\tau = c(24\tau_{MREV} + 16\tau_{\pi/2})$, where $2\tau$ represents, as before, the temporal window between the $\pi$ pulses of the $CPMG1$ sequence. As shown in Fig. 6, by applying the $CPMG1 - MREV16$ for $\tau_{MREV} = 100\mu s$; $c = 4$ and so $\tau = 9.92$ ms ($2\tau \approx 20$ ms is longer than $T_{2HE}$) we observed long tails in the magnetization, even longer than those obtained for the $CPMG1$ sequence without $MREV16$. The magnetization manifested two different decay times, a short one that could be fixed as $t_{s_{MREV}} = T_{2HE}^{MREV} \approx 0.3$ s and a longer one of $t_{l_{MREV}} = 5.3$ s. This is almost three orders of magnitude longer than $T_{2HE} \approx 15$ ms, as was also reported for silicon when this sequence was applied.[3].

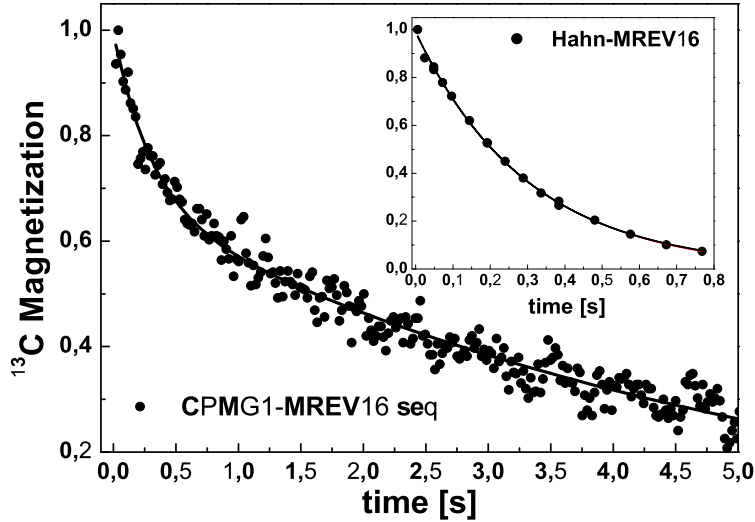

FIG. 6: Long tail obtained through the $CPMG1 - MREV16$ sequence for $\tau_{MREV} = 0.1$ ms and $c = 4$. The curve is fitted with a double exponential decay, the shorter characteristic time is fixed as $t_{s_{MREV}}$, and the longer decay is $t_{l_{MREV}} = 5.3$ s. In the inset the $Hahn - MREV16$ experiment fits an exponential decay with $T_{2MREV} = 0.3$ s.

Evidently, these long tails do not yield decoherence times, but in analogy with the situation without decoupling, they are a consequence of the stimulated echoes. Indeed, as in the pure $CPMG1$ sequence, we were able to identify a short decay time which agrees with $T_{2HE}^{MREV}$ and a long one in the same scale of times than the stimulated echo decay. Moreover, we repeated the stimulated echo decay time experiment by applying the $MREV16$ sequence between the intermediate $\pi/2$ pulses and varying $\tau_{MREV}$ between $0,1$ ms and $1$ ms and $c$ between $1$ and $90$, as expected no difference was observed when compared with the plain $SE$ sequence (see inset in Fig.3).

## III. CONCLUSIONS

The experiments shown in this work confirm the two hypotheses made in our previous work.[4] We have proved that the $C_{60}$ line is inhomogeneously broadened. This condition is the key ingredient for the formation of the stimulated echoes. In these inhomogeneous samples, the differences in frequencies from site to site are larger than the dipolar couplings, making ineffective the flip flop mechanism. Consequently, the rf inhomogeneity can not be averaged out through dipolar interactions, as would occur with homogeneous samples. The lines of the three samples (Si, $C_{60}$, $Y_2O_3$) which have manifested the presence of stimulated echoes producing long tails, have a FWHM at least one order of magnitude larger than the dipolar coupling $k = \frac{\mu_0}{4\pi} \frac{\gamma^2 \hbar}{2\pi d^3}$.

It was also verified that the decay time obtained through a Hahn echo experiment, $T_{2HE}$, is the time which characterizes the flip flop process while the long tails are pseudocoherences, a consequence of the stimulated echoes.

Another remarkable fact is the importance of the absence of the flip flop in the formation of the long tails, observed for the $CPMG1$ and $CP2$ sequences: the more time we let the flip-flop process to become effective, the shorter are the characteristic times of the long tails, as can be seen in Fig.5. These experiments are a very strong demonstration that *the long tails are a manifestation of the absence of spin diffusion.*[4]

Our explanation contrasts with the argument exposed by Li et al.[5] They claim that the delta pulse approximation is not valid and the long tails are a consequence of dipolar interactions during the pulses. Indeed, the theoretical model proposed by Li et al. to reproduce qualitatively the experimental results can be explained with the stimulated echo picture. By considering the finite duration of the pulses evolving under an artificially increased dipolar interaction one spin flip, two spin-flips and three spin-flip terms appear that can account for the phenomenological "difference in the tilting angle" with respect to a perfect $\pi$ pulse. Thus, they build a distribution of different pulses in different sites of the sample, and conse-quently, the formation of stimulated echoes with magnetization stored in the $H_0$ axis. But still, the long tail formation requires the absence of spin diffusion. Clearly, a strong flip flop mechanism during the interpulse delay will be efficient to average out the differences of the tilting angles quenching the stimulated echoes.

With regard to the effect of the finite duration of the pulses, as can be seen in Fig. (13) of ref.[19] *the $CPMG1$ tail height is insensitive to large changes in pulse strength*. In the mentioned experiment the authors obtained the same tail height for pulses strong enough to excite from 4 to 450 times the FWHM so, as all the pulses set in $C_{60}$ or in $^{29}$Si excite more than 200 or 300 times the FWHM, the delta pulse approximation is correct and the long tails can not be a consequence of finite pulse durations. Moreover, from their own average Hamiltonian calculations one can see (eqs. (4) and (5) in[5]) that as soon as the interpulse time $\tau$ is longer than the pulse duration $t_p$, the latter will have a negligible effect. This is effectively observed in their experiments in $^{29}$Si as a function of $t_p$ for $\tau = 1.096$ ms, where as soon as $t_p < 300$ $\mu$s there is no dependence on $t_p$. Everything points to the fact that the important ingredient for the long tails is a chemical shift distribution or equivalent.

From the stimulated echo decay sequence developed here, it was evident that its decay time is in the order of seconds as the long tails are. Moreover, it was proved that the characteristic SE decay is independent on the dipolar interaction: i.e. it is the same regardless the application of the $MREV16$ sequence which averages out the dipolar interaction. As a consequence, after the constructive interference between the normal and the stimulated echoes, long tails in the decay of the magnetization are found not only for the $CPMG1$ and $CP2$ sequences but also for the $CPMG1 - MREV16$.

While these results are quite disappointing with regard to the expectation created around the allegedly very long coherence times,[1–3,5] they state the basis to understand interferences which can be exploited in quantum and classical devices. As an example, the design of sequences that control the chemical shift distribution at will, possibly through gradients, can be used to preserve coherences as polarization along $z$.

We acknowledge support from Fundación Antorchas, CONICET, FoNCyT, SeCyT-UNC and the ECOS-SUD collaboration grant. We benefitted from fruitful discussions with H.M. Pastawski on localization phenomena.

* Electronic address: patricia@famaf.unc.edu.ar
[1] A. E. Dementyev, D. Li, K. MacLean, and S. E. Barrett, Phys. Rev. B **68**, 153302 (2003).
[2] S. Watanabe and S. Sasaki, Jpn. J. Appl. Phys. **42**, L1350 (2003).
[3] T. D. Ladd, D. Maryenko, Y. Yamamoto, E. Abe, and K. M. Itoh, Phys. Rev. B **71**, 014401 (2005).
[4] M. B. Franzoni and P. R. Levstein, Phys. Rev. B **72**, 235410 (2005).
[5] D. Li, A. Dementyev, Y. Dong, R. G. Ramos, and S. E. Barrett, Phys. Rev. Lett. **98**, 190401 (2007).
[6] E. Abe, K. M. Itoh, J. Isoya, and S. Yamasaki, Phys. Rev. B **70**, 033204 (2004).
[7] R. de Sousa, N. Shenvi, and K. B. Whaley, Phys. Rev. B **72**, 045330 (2005).
[8] H. Y. Carr and E. M. Purcell, Phys. Rev. **94**, 630 (1954).
[9] C. Slichter, *Principles of magnetic resonance* (Springer-Verlag, 1990), 3rd ed.
[10] E. Fukushima and S. Roeder, *Experimental pulse NMR, A nuts and bolts approach* (Addison Wesley, 1981).
[11] D. P. DiVincenzo, Science **270**, 255 (1995).
[12] B. Kane, Nature **393**, 133 (1998).
[13] D. G. Cory, A. F. Fahmy, and T. F. Havel, Proc. Natl. Acad. Sci. U.S.A. **94**, 1634 (1997).
[14] E. L. Hahn, Phys. Rev. **80**, 580 (1950).
[15] S. Zhang, B. H. Meier, and R. R. Ernst, Phys. Rev. Lett. **69**, 2149 (1992).
[16] P. R. Levstein, G. Usaj, and H. M. Pastawski, J. Chem. Phys. **108**, 2718 (1998).
[17] G. Usaj, H. M. Pastawski, and P. R. Levstein, Mol. Phys. **95**, 1229 (1998).
[18] H. M. Pastawski, P. R. Levstein, G. Usaj, J. Raya, and J. Hirschinger, Physica A **283**, 166 (2000).
[19] D. Li, Y. Dong, R. G. Ramos, J. D. Murray, K. MacLean, A. E. Dementyev, and S. E. Barrett, ArXiv:0704.3620 (2007).
[20] P. L. Kuhns and M. S. Conradi, J. Chem. Phys. **77** (1982).
[21] W. K. Rhim, D. D. Elleman, and R. Vaughan, J. Chem. Phys. **59**, 3740 (1973).
[22] P. Mansfield, M. J. Orchard, D. C. Stalker, and K. H. B. Richards, Phys. Rev. B **7**, 90 (1973).
[23] U. Haeberlen, *High resolution NMR in Solids: Selective Averaging - Supplement 1 for Advances*

*in Magnetic Resonance* (Academic Press, 1976).